\documentclass[manuscript, nonacm]{acmart}

\usepackage{tabularx}
\usepackage{booktabs}
\usepackage{array}
\usepackage{makecell}
\usepackage{graphicx}
\usepackage{enumitem}

\AtBeginDocument{%
  }

\setcopyright{acmlicensed}
\copyrightyear{2018}
\acmYear{2018}
\acmDOI{XXXXXXX.XXXXXXX}
\acmConference[Conference acronym 'XX]{Make sure to enter the correct
  conference title from your rights confirmation email}{June 03--05,
  2018}{Woodstock, NY}
\acmISBN{978-1-4503-XXXX-X/2018/06}

\begin{document}

\title[Exploring \textit{Safe Visibility} for Graduate Student Mothers] {``A Space Where I Can Be Myself'': Exploring \textit{Safe Visibility} for Graduate Student Mothers in Mom Cafes}
\author{Jeanne Choi}
\email{jeannechoi@u.northwestern.edu}
\affiliation{%
  \institution{Northwestern University}
  \city{Evanston}
  \country{United States}
}

\author{Joohee Choi}
\email{lizchoi01@kaist.ac.kr}
\affiliation{%
  \institution{KAIST}
  \city{Daejeon}
  \country{Republic of Korea}
}

\author{Jeong-woo Jang}
\email{jangjw29@kaist.ac.kr}
\affiliation{%
  \institution{KAIST}
  \city{Daejeon}
  \country{Republic of Korea}
}

\renewcommand{\shortauthors}{Choi et al.}

\begin{abstract}
  Graduate student mothers navigate intertwined demands of graduate study and caregiving, yet little is known about how they seek support through online communities. We examine this through Mom Cafes, locally based online parenting communities in South Korea. Drawing on focus group interviews and design workshops with 22 graduate student mothers, we identify a pattern of dual invisibility: caregiving circumstances were difficult to make legible in academic contexts, while academic circumstances were difficult to make legible in parenting contexts. Although Mom Cafes offered useful local information and emotional reassurance, concerns specific to graduate student motherhood were harder to articulate or safely disclose. Participants sought ways to make graduate student motherhood recognizable when relevant, but their designs highlighted tensions around identifiability, judgment, relational burden and accountability. We conceptualize these tensions as \textit{safe visibility} and discuss infrastructures for selective recognition, low-obligation connection, and accountable pathways to practical care.
\end{abstract}





\maketitle

\section{Introduction}
Graduate student mothers navigate simultaneous demands of graduate education, research, and caregiving~\cite{gsm07, gsm09}. In addition to being mothers, they often occupy multiple academic roles as students, researchers, instructors, and employees, whose expectations may conflict with pregnancy, childbirth, and childcare responsibilities~\cite{gsm08, gsm10}. Prior research on this community has documented persistent challenges associated with their position. Time scarcity~\cite{gsm02}, emotional strain~\cite{gsm01, gsm13}, financial precarity~\cite{gsm03, gsm04, gsm12}, and limited institutional accommodation~\cite{gsm11} are highly associated with navigating their overlapping demands.

When institutional resources and personal networks are insufficient, online parenting communities can serve as supplementary sources of support. Prior research has shown that online parenting communities act as valuable infrastructures for information exchange~\cite{progga23post, rhee2023understanding}, emotional reassurance~\cite{gui17invest}, experiential knowledge~\cite{ibrahim24understand, gibson13digital}, and social connection~\cite{welbourne09supp}. Yet for graduate student mothers, some support needs arise not from motherhood or graduate study alone, but from the interaction between the two. Online parenting communities are typically organized around a broad shared identity such as parenthood or motherhood, even though their members simultaneously occupy diverse social positions. Consequently, the support a person needs may extend beyond the identity that provides the primary basis for participation in the community. Prior research has examined these circumstances by investigating online communities created for particular populations such as low-income students~\cite{pyle23social} and LGBTQ individuals~\cite{dym19coming}, or for parents situated at particular intersections of identities and social positions, including race, gender identity, and migration status~\cite{bauer22mom, robinson2019takes, blackwell2016lgbt, veazey2019glocalised}. We know less, however, about how people seek support in broad online communities when their needs depend on circumstances beyond the shared identity that brings the community together. In such settings, the question is not simply whether members can disclose another aspect of themselves, but how the community’s organizing identity shapes which experiences appear relevant, understandable, and appropriate for support-seeking.

We examine this question in the context of \textit{Mom Cafes}, locally based online parenting communities widely used by mothers in South Korea to exchange childcare information and social support. Because these communities are organized broadly around motherhood rather than specific circumstances such as graduate student motherhood, they provide a useful setting for examining when different aspects of users’ circumstances become relevant to their support needs. Thus in this study, we investigate how graduate student mothers in South Korea seek and navigate online support in Mom Cafes while balancing graduate study and motherhood, as well as what forms of support they envision as better addressing their needs. We ask:

\begin{itemize}
    \item RQ1. How do graduate student mothers seek and navigate support through Mom Cafes while balancing graduate study and motherhood?
    \item RQ2. What forms of online support do graduate student mothers envision as better addressing their needs, and what tensions arise in designing such support?
\end{itemize}

To address these questions, we conducted focus group interviews and participatory design workshops with 22 graduate student mothers in South Korea. The interviews explored participants’ broader experiences of graduate student motherhood, their use and non-use of Mom Cafes and other online communities, and the forms of support they sought across these spaces. In the subsequent workshops, participants envisioned online communities that could better support graduate student mothers. Connecting participants’ accounts of existing support practices with their designs for alternative forms of support allowed us to examine both the limitations of current online support infrastructures and the conditions participants considered necessary for improving them.

Our findings show that participants encountered a pattern of \textit{dual invisibility} across academic and parenting contexts: academic contexts often overlooked their caregiving circumstances, while parenting contexts left little room for their academic identities. This fragmented recognition also shaped their support-seeking pattern in online spaces. Participants found Mom Cafes highly valuable for parenting support, yet concerns arising specifically from the overlap of graduate study and motherhood were often harder to articulate, less readily understood, or riskier to disclose. However, participants further showed that merely increasing their visibility as graduate student mothers alone was not sufficient. Participants sought connection with similar others while remaining wary of exposure, judgment, relational obligations, and accountability. We conceptualize this set of tensions as \textit{safe visibility}: becoming recognizable to relevant others for a particular support need while retaining control over which aspects of oneself become visible, how identifiable one becomes, and what relational demands follow from that visibility.

Through this work, we make three contributions. First, we provide an empirical account of how graduate student mothers seek and assemble online support while navigating the intertwined demands of graduate study and caregiving. Second, we conceptualize safe visibility as the tension between becoming recognizable for support and managing identifiability and relational demands. Third, we offer design considerations for enabling relevant connections while preserving control over identity, obligation, and accountability. Together, these contributions highlight the need for online communities to support selective, need-based visibility rather than uniform identity disclosure.

\section{Related Work}
We situate our work across three areas of prior research on the challenges faced by graduate student mothers, online parenting communities as social support infrastructures, and identity and online support. Together, these bodies of work provide a foundation for examining how support needs that span academic and caregiving domains are navigated across online spaces organized around different shared identities.

\subsection{Challenges of Graduate Student Mothers}
Although the transition to motherhood presents substantial challenges in itself~\cite{Role23Pai}, graduate student mothers ~---~ those who pursue graduate education while raising children ~---~ encounter additional demands arising from the multiple roles they occupy. In addition to being mothers, they often serve as students, researchers, instructors, and employees. Because these roles involve distinct and sometimes conflicting responsibilities and expectations~\cite{gsm08, gsm10, gsm06}, graduate student mothers must continually balance across overlapping domains of academic and parenting~\cite{gsm07, gsm09}. As these responsibilities accumulate, they frequently experience chronic time scarcity, making it difficult to meet both academic obligations and childcare responsibilities~\cite{gsm02}. These pressures can further lead to emotional and psychological distress, particularly when they feel unable to devote sufficient time and attention to either their academic work or their children~\cite{gsm01, gsm13}. Within South Korea, prior qualitative work has similarly documented the challenges that arise from the role conflict of graduate student mothers~\cite{seo15gsm}.

Many of these challenges resemble those of employed mothers, who likewise navigate competing demands of paid work and caregiving~\cite{westrupp15work, lee17work}. Yet graduate student mothers face challenges specific to their position within academia. Universities often provide limited or inconsistent accommodations for student parents~\cite{gsm11}, while institutional provisions that could substantially alleviate parenting burdens ~---~ such as childcare subsidies, evening and weekend childcare, financial assistance, and paid parental leave ~---~ are frequently inadequate or unavailable~\cite{gsm05, gsm06}. They may also experience financial precarity because they are still in the process of establishing their careers~\cite{gsm04} and often depend on comparatively limited stipends or other forms of temporary funding~\cite{gsm03, gsm12}. These conditions can intensify the difficulty of reconciling academic progress with caregiving responsibilities, highlighting the importance of examining graduate student mothers as a population with distinct support needs.

Much of this prior research on graduate student mothers has examined how they navigate their challenges within academic institutions and their immediate social environments. Yet when institutional support falls short, parents frequently turn to online communities for social support~\cite{parenting01, parenting02}. Also, more recent research has begun to examine how student mothers use digital technologies to navigate parenting and academic life~\cite{bhatti21parenting, bhatti22intimate, gsm12}. This raises the need to understand how graduate student mothers seek support through broadly organized online parenting communities when their support needs arise specifically from the intersection of graduate study and motherhood.

\subsection{Online Parenting Communities as Social Support Infrastructures}
Online communities have become important venues for exchanging social support~\cite{online01, andalibi18social, pyle23social, dym19coming, razi20lets, huh23help}, particularly around health concerns and major life transitions~\cite{zhang22separate, corvite22social, yang19channel, yang19seekers}. In addition to supplementing offline support networks, they offer features such as asynchronous communication~\cite{tanis2008health} and access to geographically dispersed others with relevant experiences~\cite{young19fire}. Parents likewise turn to online communities for support in navigating pregnancy, childbirth, and caregiving. Prior research has shown that online parenting communities serve as important sources of parenting information, practical advice, and experiential knowledge~\cite{ibrahim24understand, gibson13digital}, while also providing opportunities for emotional reassurance and social connection~\cite{gui17invest}.

Meanwhile, a substantial body of research has focused specifically on online communities for mothers, whose support needs may be shaped by physical, emotional, and social changes associated with pregnancy, childbirth, and caregiving~\cite{parry13from, gui17invest, yadav22clinic, newhouse16bump}. For example, studies of online communities for women experiencing postpartum depression have found that these spaces can facilitate disclosure of emotional distress, exchange of coping strategies, and reassurance from others with similar experiences~\cite{progga23post, rhee2023understanding, evans12social}. Research on online infertility support groups has likewise shown how interaction with others facing comparable circumstances can foster a sense of virtual community and reduce feelings of isolation~\cite{welbourne09supp}. Across these settings, online communities can enable parents to access not only information but also experiential knowledge that helps them interpret their own circumstances through the experiences of others.

At the same time, the support available through these communities is shaped by the experiences around which interaction is organized. Prior work has demonstrated the value of connecting with others who share relevant circumstances, as such similarity can shape processes of social support and mutual understanding~\cite{parenting05, hargreaves18share}. Yet parents who participate in a broadly organized parenting community may differ substantially in the occupational, educational, familial, and social circumstances that shape particular needs. This raises a broader question about what happens when the support a member requires is not fully captured by the shared identity around which a community is organized. In our research, we examine this gap by investigating the experience of graduate student mothers in Mom Cafes.

\subsection{When Identity Matters for Online Support}
People participate in online communities while occupying multiple social identities and circumstances, whose relevance may shift depending on the situations they encounter and the support they seek~\cite{farnham11facet}. HCI scholarship has challenged approaches that treat social categories as isolated or additive, emphasizing instead how people’s experiences emerge through relations among multiple aspects of their social positions and the contexts in which technology use takes place~\cite{sch17inter}. This perspective is particularly relevant to parenting, as experiences of motherhood are shaped by women’s broader social positions and life circumstances. Prior research has examined online spaces for mothers situated in particular circumstances, including academic mothers~\cite{bauer22mom, bhatti21parenting, bhatti22intimate}, African American mothers~\cite{robinson2019takes}, LGBT parents~\cite{blackwell2016lgbt, Alang02012015}, and migrant parents~\cite{veazey2019glocalised}. These studies show how connecting with others who share relevant experiences can facilitate mutual understanding, disclosure, solidarity, and access to contextually relevant support. Much of this work, however, has focused on spaces organized around a particular population or shared circumstance. In these spaces, members’ relevant experiences are already relatively recognizable to others. Broader communities organized around a shared identity pose a different challenge. Although members share the identity that defines the community, other aspects of their circumstances may become important for particular support needs. The community’s organizing identity can therefore shape which experiences are readily understood as relevant and appropriate for support-seeking.

When a support need cannot be understood through the community’s shared identity alone, members may need to make other aspects of their circumstances visible. Here, research on identity management and disclosure in online communities shows that such visibility is itself negotiated. People manage what they reveal, to whom, and in which contexts, while affordances such as anonymity and pseudonymity can enable disclosure while limiting unwanted exposure~\cite{ma16anon, andal16sexual, kang13why}. People may also separate identities and networks across online spaces when particular experiences are difficult or risky to discuss within their existing networks~\cite{haimson18social, zhang22separate}. Prior work has therefore highlighted tensions between connection and disclosure on the one hand and privacy and control on the other~\cite{lamp11we}. In broader communities, however, these tensions concern not only how much of oneself to disclose, but also whether and when a particular aspect of one’s circumstances needs to become recognizable for a support. Doing so may facilitate understanding and connection with relevant others, while also introducing risks of exposure, identifiability, judgment, or unwanted relational involvement.

This tension is particularly relevant to graduate student mothers participating in Mom Cafes, which is a general online community for mothers. Graduate study need not be salient when members seek parenting information or reassurance that can be understood through their shared experience as mothers. However, it may become consequential when a support need arises specifically from the overlap of academic and caregiving responsibilities. We therefore examine how graduate student mothers navigate Mom Cafes when their academic circumstances become relevant to support-seeking, and what tensions emerge when designing for need-based recognition, connection, and practical support.

\section{Method}

Our study aimed to explore how graduate student mothers experience and navigate Mom Cafes, as well as how such communities might better support their needs. To examine both participants’ lived experiences and their unmet needs for future support, we conducted an integrated study session combining a focus group interview with a design workshop. Although all participants were current graduate students, many of the discussions centered on retrospective experiences of pregnancy, childbirth, childcare, and online community use while balancing graduate study. We therefore employed a focus group interview to facilitate collective reflection by allowing participants to recall and elaborate on past experiences through one another's narratives~\cite{acocella2012focus, morrison1998search}. To complement participants’ accounts of existing practices, we subsequently conducted a design workshop to elicit unmet needs and explore how online communities might better support graduate student mothers. This sequence allowed participants to draw on experiences discussed in the focus group when generating design ideas in the subsequent workshop. More broadly, design workshops have been used in HCI to uncover latent user needs and generate design implications for sociotechnical systems~\cite{choi23toward, kim23we, lee22distract}. This study was approved by our institution’s Institutional Review Board (IRB).

\subsection{Study Context: Mom Cafes}
Mom Cafes are online communities for mothers in South Korea, many of which are organized around particular residential areas and facilitate the exchange of locally relevant information and discussion~\cite{bae15mom, momcafe1}. Prior studies have documented their use for seeking and exchanging information related to pregnancy, childbirth, childcare, education, health, and everyday life, as well as for connecting through shared experiences of motherhood~\cite{momcafe2, wang21mother, hong12smart}. Although individual Mom Cafes vary in size, rules, and internal organization, they are structured around topic-specific boards and public posts~\cite{momcafe1}. Thus Mom Cafes are not only information repositories but also locally embedded social spaces shaped by ongoing participation. These features, including locality, public posting, topic-based organization, and participation requirements, were salient characteristics of the study context and informed our attention to how participants described seeking support.

\subsection{Participants}
\label{sec:participant}

\begin{table}[t]
\centering
\caption{Demographic information of user study participants. Twenty-two participants were recruited to reflect variation in academic background, length of experience balancing graduate study and parenting, and Mom Cafe use.}
\Description{A six-column table with 22 participant rows, listing each participant’s ID, academic major, degree, length of experience balancing graduate study and parenting, number of Mom Cafe visits, and other online communities used. Participants included both master's and PhD students across diverse disciplines, with widely varying levels of Mom Cafe activity and use of platforms such as Kakao Open Chat, Instagram, Danggeun, and ChatGPT.}
\label{tab:participants}
\small
\begin{tabularx}{\textwidth}{@{} l >{\raggedright\arraybackslash}X l >{\raggedright\arraybackslash}X >{\centering\arraybackslash}p{1.7cm} >{\raggedright\arraybackslash}X @{}}
\toprule
\textbf{ID} 
& \textbf{Major} 
& \textbf{Degree} 
& \makecell{\textbf{Length of Balancing}\\\textbf{Studies and Parenting}} 
& \makecell{\textbf{Mom Cafe}\\\textbf{Visits}\textsuperscript{*}} 
& \makecell{\textbf{Other Used Online}\\\textbf{Communities}} \\
\midrule
P1  & Korean Education               & Master's & 1 year and 4 months  & 2635  & Instagram \\
P2  & Language                       & PhD      & 1 year and 7 months  & 1837  & - \\
P3  & Math Education                 & PhD      & 5 years and 4 months & 1016  & Instagram \\
P4  & Medical Humanities             & PhD      & 1 year and 6 months  & 2797  & Kakao Open Chat \\
P5  & Speech-Language Pathology      & Master's & 2 months             & 5106  & Naver Blog, Instagram \\
P6  & Social Welfare                 & PhD      & 7 years              & 11432 & Kakao Open Chat, Instagram, Danggeun \\
P7  & Law                            & Master's & 7 months             & 578   & - \\
P8  & Social Welfare                 & PhD      & 5 years and 2 months & 6000  & Kakao Open Chat \\
P9  & Sociology                      & PhD      & 2 years and 6 months & 3766  & Kakao Open Chat \\
P10 & Social Studies                 & Master's & 2 months             & 1668  & Kakao Open Chat, Instagram, Danggeun \\
P11 & Education                      & PhD      & 8 months             & 1705  & Instagram \\
P12 & Korean Language and Literature & PhD      & 2 years              & 525   & Danggeun, Instagram, Facebook, Naver Band, Thread \\
P13 & Education                      & PhD      & 5 years              & 1023  & Kakao Open Chat, Danggeun \\
P14 & Counseling Psychology          & Master's & 2 years and 6 months & -     & Danggeun, Youtube, Instagram \\
P15 & Education                      & PhD      & 4 years              & -     & - \\
P16 & Biochemical Engineering        & PhD      & 5 months             & 26    & Mommy Talk, ChatGPT, Youtube \\
P17 & Communication                  & PhD      & 1 year and 6 months  & 35    & Danggeun, Mommy Talk, ChatGPT \\
P18 & Computer Science               & PhD      & 4 years and 6 months & 50    & - \\
P19 & International Politics         & PhD      & 3 years              & 70    & Kakao Open Chat, ChatGPT \\
P20 & Science and Technology Policy  & PhD      & 1 year and 10 months & 840   & Kakao Open Chat, Instagram, Thread, ChatGPT \\
P21 & Public Administration          & PhD      & 2 years and 6 months & 59    & Mommy Talk, Instagram, Thread \\
P22 & Early Childhood Education      & PhD      & 5 years              & -     & ChatGPT, Instagram \\
\bottomrule
\addlinespace[2pt]
\multicolumn{6}{@{}p{\textwidth}@{}}{\footnotesize \textsuperscript{*}Mom Cafe Visits refers to the number of visits displayed by the platform for each participant’s account and reported by the participant in the pre-survey. A dash indicates no reported Mom Cafe use.}
\end{tabularx}
\end{table}

We recruited twenty-two participants through multiple online and personal networks. Eligible participants were women who were currently enrolled in a graduate program and had experience balancing graduate studies with pregnancy, childbirth, and/or childcare. Prior experience with Mom Cafes was not required because we sought to understand not only how graduate student mothers use these communities, but also why some discontinue or never adopt them. To ensure variation in experience with Mom Cafes, participants were recruited through both Mom Cafes and independent channels, including local online communities, online messaging groups for university-affiliated kindergartens, and snowball sampling.

We first asked interested participants to complete a pre-survey assessing their current and previous use of Mom Cafes, as well as their use of other online communities as graduate student mothers. For participants with prior Mom Cafe use, the pre-survey also asked them to report the number of visits displayed by the platform for their account at the time of survey completion. The pre-survey also collected demographic and background information, including age, major, region of residence, stage of graduate study, number and age of children, and the length of time they had balanced graduate studies and parenting. We used this information to construct diverse groups that represented a broad range of experiences with graduate student motherhood, parenting, and online community use (Table~\ref{tab:participants}).

From the pool of respondents, we purposively selected twenty-two participants to maximize variation in Mom Cafe use, academic backgrounds, and experiences of balancing graduate study and parenting: thirteen current Mom Cafe users and nine former or non-users. We defined current Mom Cafe users as participants who reported currently using Mom Cafes (e.g., visiting, reading, posting, or commenting), while former or non-users included both participants who had used them previously but no longer did and those who had never used Mom Cafes. By including users with various use experiences, we sought to capture not only how graduate student mothers use Mom Cafes but also why some discontinue or avoid using them rather than treating them as predefined comparative groups.

Based on the pre-survey responses, we organized the selected participants into five group sessions. To facilitate discussion among participants with broadly similar experiences of Mom Cafes while maintaining diversity within each group, we grouped participants by their current Mom Cafe use status and varied group composition by academic major and the length of time balancing graduate studies and childcare. The current Mom Cafe users were assigned to three groups (two groups of four and one group of five), while former or non-users were assigned to two groups (one group of four and one group of five).

\subsection{Study Procedure}
Each session lasted approximately two hours, consisting of a one-hour focus group interview, a 10-minute break, and a 50-minute design workshop. All sessions were conducted remotely via Zoom because participants were enrolled in universities across different regions. 

\subsubsection{Focus Group Interview}
The first part of the session was a semi-structured focus group interview. At the beginning of the interview, the researcher explained how the group discussion would proceed. Participants generally responded in turn, while those with relevant experiences were invited to elaborate voluntarily and build on others' responses. Interview questions were displayed on shared slides throughout the session to support discussion.


The interview covered participants’ academic and parenting backgrounds, challenges of graduate student motherhood, offline support, and online community use. Current Mom Cafe users were asked about their motivations for continued use, how Mom Cafes supported them as mothers or graduate student mothers, and negative or limiting experiences. To encourage detailed reflection, current users were invited to refer to their own Mom Cafe activity, such as previous posts or comments. Former or non-users of Mom Cafes were asked why they had never used or had stopped using these communities, as well as the experiences and perceptions that affected those decisions. Both groups also discussed other online communities they used in relation to graduate student motherhood and their motivations behind those choices.


\subsubsection{Design Workshop}
Following the focus group interview and a short break, participants took part in a design workshop. Instead of asking participants to imagine an ideal online community in the abstract, we first asked them to identify concrete challenging situations they had experienced as graduate student mothers and then use those situations as starting points for design ideation. Therefore the workshop consisted of two activities, and we used a shared Google Slides file as the collaborative workspace. In \textit{Part 1}, participants individually described specific challenging situations they had experienced as graduate student mothers. We encouraged participants to describe concrete episodes involving competing academic and caregiving demands rather than general difficulties. Participants wrote their scenarios on the shared slide, allowing everyone to view one another’s responses and use them as prompts for further reflection and idea generation.

In \textit{Part 2}, participants were asked to sketch a Mom Cafe that would have helped address one or more of the scenarios generated in \textit{Part 1}. We retained the term ``Mom Cafe'' to ground the design task in the type of locally embedded parenting community discussed during the focus group, while describing it in broad functional terms as a local online community for exchanging parenting information and support. This framing allowed participants to draw on their familiarity with Mom Cafes without constraining their designs to the features or structures of existing platforms. Participants created individual sketches using pen and paper, photographed their designs, and uploaded the images to the shared Google Slides file.

After uploading their sketches, participants presented their designs to the group. They explained (1) how the proposed design would support graduate student mothers, and (2) which specific challenge or unmet need it was intended to address. As participants presented their designs in turn, they were able to see and respond to ideas raised by earlier participants, sometimes relating their own designs to those presented before them. To conclude the workshop, we asked, ``What kind of online community would be most helpful for women who are balancing graduate school with pregnancy, childbirth, and parenting?'' This final question provided an opportunity for participants to collectively reflect on their unmet needs, expectations, and ideas for online communities.

\subsection{Data Analysis}
With participants’ consent, all sessions were audio- and video-recorded via Zoom. We also collected the collaborative Google Slides files generated during the design workshop, including participants’ written scenarios and uploaded sketches. We analyzed these materials together with participants’ verbal explanations and group discussions, treating the artifacts as representations of proposed support mechanisms rather than as standalone design outcomes.

We analyzed the interview and design workshop data using an iterative, codebook-oriented thematic coding approach~\cite{guest12anal, braun20ta}. During analysis, we attended to participants' Mom Cafe use status as contextual information but developed themes across the full dataset rather than conducting a group-based comparison. Audio recordings from the interviews and design workshops were transcribed using Naver Clova, a Korean speech-to-text service\footnote{https://clovanote.naver.com/}. For each group session, we combined the interview and workshop transcripts into a single analytic document. To develop the codebook, the first and second authors began by independently coding the transcript from one session. They then met to compare codes, discuss overlapping themes and discrepancies, and construct an initial codebook. Using this initial codebook, the same two authors independently coded the transcript from a second session. They subsequently met again to refine the codebook by discussing newly emergent codes and resolving coding discrepancies, resulting in a revised codebook. The first author then applied this revised codebook to the remaining three session transcripts. This iterative process produced a final codebook consisting of 121 codes.

All sessions were conducted in Korean, and the transcripts were analyzed in Korean. The first author translated quotations selected for the manuscript into English, and the second author reviewed the translations against the original Korean excerpts to preserve participants' intended meanings.
\section{Findings}
In this section, we examine how graduate student mothers negotiated the visibility of their academic and caregiving circumstances across academic, parenting, and online community contexts. Although participants faced various financial, institutional, and physical challenges as a graduate student mother, our analysis centers on a challenge that consistently emerged across both interviews and workshops: negotiating when and how academic and caregiving circumstances needed to become recognizable for particular support needs to be understood and supported. We first situate participants' Mom Cafe use within a broader pattern we characterize as \textit{dual invisibility}, where different social contexts recognized only one dimension of their identity. We then examine how this fragmented recognition shaped participants' use of Mom Cafes and how the design workshop surfaced tensions involved in creating \textit{safe visibility} in online parenting communities.

\subsection{Dual Invisibility Across Academic and Parenting Contexts}

Participants described how different contexts made different aspects of their circumstances as graduate student mothers legible, while the intersection of graduate study and caregiving was often difficult to make relevant when support depended on both. Although they simultaneously occupied the roles of graduate student, researcher, and mother, different social contexts recognized only one part of this intersecting identity. Across both academic and parenting environments, participants therefore continually negotiated which aspects of themselves became visible depending on the context. In the following subsections, we describe how this fragmented visibility emerged, and how it led participants to selectively disclose or withhold parts of themselves.

\subsubsection{Motherhood is Invisible in Academic Spaces}
Participants discussed the difficulty of being understood as mothers within their academic and research environments, including their research groups and relationships with advisors and peers. Many were surrounded by peers who were single or did not have children, making it difficult to discuss pregnancy or childcare-related struggles in ways others could readily understand. Participants further shared that advisors and lab members also did not always accommodate these demands, even when they were aware of them. For example, P2 recalled being asked to lead a campus tour for visitors despite experiencing severe morning sickness. Participants thus often felt they had to repeatedly \textit{explain} their situations before asking for understanding or support, which itself was emotionally burdensome.

\begin{quote}
    \textit{``In my field, I am the only woman in my lab, and the other students are all men. Most of them are unmarried, in their twenties, and do not have children. So there were issues they simply could not empathize with because we were at different life stages. I always had to explain my situation and ask for their understanding, which was psychologically difficult.''} (P18)
\end{quote}

The absence of shared experience also extended beyond individual labs. Even when participants sought peers who had navigated similar circumstances, such as conducting laboratory work during pregnancy, they rarely found others whose academic and parenting timelines aligned. These conditions limited access to emotional and informational support within their academic institutions.

As a result, participants were often selective about when and how they discussed their responsibilities as mothers. Rather than frequently raising the constraints associated with pregnancy, childbirth, or childcare, they sometimes chose to keep these experiences in the background to maintain their roles as students and researchers.

\subsubsection{Graduate Study is Invisible in Parenting Contexts}
\label{sec:parenting}

Participants experienced a similar lack of recognition in parenting contexts. Although they could connect with other mothers through shared experiences of motherhood, they struggled to have their identities as graduate students recognized. Graduate student mothers were rare in their parenting networks, leaving few opportunities to seek advice or share experiences related to academic work. As in academic settings, graduate student motherhood was often an unfamiliar identity to explain, leading participants to remain silent.



Participants frequently contrasted their situations with those of \textit{working moms}, a commonly used term in South Korea for mothers who engage in paid work while raising children. While both groups manage multiple responsibilities, participants perceived working mothers' challenges as more readily understood. In contrast, graduate study was often perceived as a personal choice or form of self-development, making it harder to justify academic commitments when they conflicted with childcare responsibilities.

\begin{quote}
    \textit{``When people say `working mom,' I think there is now more social awareness that working moms deserve empathy and consideration. But a mother who is studying is different, because studying is seen less as a job and more as personal growth or self-development. Work and childcare are at least recognized because work is tied to making a living. But when studying and childcare are placed side by side, prioritizing studying can make people see me as a selfish mother.''} (P8)
\end{quote}

Some participants also encountered discomfort when disclosing their educational backgrounds in parenting communities. Parenting contexts thus offered support for issues related to motherhood while leaving their experiences as graduate students unacknowledged.




\subsubsection{Fragmented Recognition Across Academic and Parenting Contexts}

We refer to this fragmented recognition as \textit{dual invisibility}: participants were visible as graduate students in academic settings and as mothers in parenting settings, yet rarely recognized as both. They therefore selectively foregrounded or withheld parts of themselves according to what each context made socially legible. For example, P8 told other mothers that she worked part-time rather than explaining that she was pursuing graduate study.



\begin{quote}
    \textit{``Being a graduate student, especially a full-time graduate student, can feel somewhat ambiguous to reveal. In my case, because I spend most of the day on childcare and housework, I naturally tend to recognize or introduce myself first as a housewife. (···) So these days, when I introduce myself to people I meet for the first time, I usually say first that I am a housewife. My identity as a graduate student comes later, or sometimes I skip it.''} (P17)
\end{quote}

The burden of dual invisibility produced not only practical strain but also emotional isolation. Participants had to continually manage when, where, and to whom each part of their identity and the related challenges such as guilt, career anxiety, and the tension between academic and caregiving expectations could be safely disclosed.

\begin{quote}
    \textit{``It felt different from being a working mom because being a graduate student is not a stable status. It was also completely different from being a housewife because the rhythm of life is so different. I felt like I did not fully belong anywhere. (···) Because of that, it often felt like a lonely struggle.''} (P21)
\end{quote}

Importantly, the difficulty did not arise simply because every context recognized only part of who participants were. Dual invisibility became consequential when the unrecognized part of their circumstances was necessary to make a need understandable or supportable ~---~ for example, when childcare disrupted research or when academic commitments conflicted with expectations of motherhood. This distinction provides important context for participants’ use of Mom Cafes, where recognition as mothers often enabled meaningful parenting support but became insufficient when their needs arose specifically from the interaction of graduate study and caregiving.

\subsection{Mom Cafes: Useful as Mothers, Invisible as Graduate Students}
In this section, we show how participants benefited from Mom Cafes as mothers while remaining less visible as graduate students, and how this uneven visibility shaped what they sought, disclosed, and received in these communities.

\subsubsection{Mom Cafes Function as Useful Parenting Infrastructure}
 
Participants who actively used Mom Cafes described them as important sources of practical parenting information. Many began using Mom Cafes when they lacked offline sources of advice or when they preferred not to directly ask acquaintances about pregnancy, childbirth, or childcare-related concerns. Because posts were organized around topics such as pregnancy stage, child age, childcare facilities, and local neighborhoods, participants could efficiently search for information that was immediately relevant to their everyday parenting needs. The region-based structure of Mom Cafes was particularly valuable, as it allowed participants to access practical and localized information about hospitals, daycare centers, kindergartens, after-school programs, and other childcare-related resources in their area.

Beyond informational support, participants also used Mom Cafes as repositories of shared parenting experiences. Reading archived posts helped them compare their own situations with those of other mothers and reduced feelings of isolation. For example, P2 described searching for posts about pregnancy symptoms by week whenever she experienced a new physical symptom. Seeing that other mothers had gone through similar situations reassured her that her experiences were not unusual. Similarly, P8 noted that reading other mothers' difficult experiences helped her feel that she was \textit{``not the only one.''}

Although many participants described themselves as relatively passive users who primarily searched for information or read archived posts, some also posted questions or comments when they needed immediate advice or emotional reassurance. Participants explained that Mom Cafes were useful precisely because many members were active and willing to respond. In some cases, posting in a Mom Cafe helped participants receive practical advice as well as empathy from other mothers, especially from those who were also balancing childcare with work-like responsibilities.

\begin{quote}
    \textit{``Even though the nature of the work is different for graduate students and working moms, they are people who are working in some way. So on days when I was having a really hard time, or when I was feeling psychologically overwhelmed, I would sometimes write posts almost as a way of venting. And when people replied, saying that they understood how I felt, it was comforting. It helped me feel a little less frustrated and eased my mind.''} (P2)
\end{quote}

Mom Cafes thus functioned as accessible parenting infrastructures that combined informational and emotional support.





\subsubsection{Graduate Student Motherhood Remains Largely Invisible in Mom Cafes}
Despite their usefulness, participants did not perceive Mom Cafes as spaces where they could comfortably bring their graduate-study circumstances into view when those circumstances shaped a support need. While the interactions in Mom Cafes were organized around a broadly shared identity as mothers, participants' experiences were also shaped by academic responsibilities. 

One barrier was the perceived lack of similar others. Participants repeatedly noted that graduate student mothers appeared to be a very small minority within Mom Cafes. Although Mom Cafes included diverse groups of mothers, participants rarely encountered posts or boards specifically addressing graduate student motherhood. During the interview, P10 searched the Mom Cafe she used and could not find relevant posts. P11 similarly noted that while there were boards for working mothers or mothers of children at particular schooling stages, she could not find a space where graduate student mothers could gather around shared concerns. Because participants saw few similar cases, they anticipated that sharing their experiences would require extensive explanation and justification, echoing the burden they had experienced in offline parenting contexts. P12, for instance, described Mom Cafes as spaces where she did not necessarily need to reveal identities such as being a graduate student mother. Rather than viewing Mom Cafes as spaces for presenting the full self, several participants treated them as informational sources for selectively accessing parenting-related support.

 


A second barrier was the risks that participants anticipated when disclosing their identity as a graduate student. Extending the perceptions described in Section~\ref{sec:parenting}, participants expected other mothers to read graduate study as optional and to ask why they continued studying while raising children. P10, for instance, assumed such reactions and had therefore never considered raising these concerns in a Mom Cafe at all.

\begin{quote}
    \textit{``Here, it somehow felt like working moms are understood as being in situations where they have no choice but to work, but being a graduate student feels a bit more optional. I think I had this assumption that people here wouldn’t really understand---like, `If parenting is so hard, why don’t you just focus on parenting? Why are you raising a child while also being in graduate school?' So I don’t think I had ever thought about talking in a Mom Cafe about what I go through as a graduate-student mother.''} (P10)
\end{quote}

This concern was not only hypothetical. Some participants had directly encountered negative reactions when graduate study was discussed in Mom Cafes. P8 described receiving harsh criticism after sharing her struggles as a graduate student mother, including comments that framed her as \textit{``a selfish mother for not focusing on childcare.''} Witnessing or directly receiving such comments reinforced participants' concerns that disclosing their identities as graduate students could invite moral evaluation of their motherhood. In particular, accounts from former and non-users illustrated how these anticipated irrelevance or judgment could lead participants to avoid or discontinue Mom Cafe use.


Participants also worried that disclosing details about their academic lives could make them identifiable. The locality-based structure that made Mom Cafes useful also made some participants feel less anonymous. Because members often lived in the same region and shared local information, participants worried that disclosing their specific academic field could make them recognizable. P3, who lived in an area associated with a particular occupation, explained that members seemed especially cautious of sharing personal stories because the space did not feel completely anonymous. Several participants therefore avoided discussing challenges specific to graduate student motherhood because explaining those challenges often required disclosing detailed information about their academic circumstances. 

Because Mom Cafes did not reliably support needs that depended on participants’ graduate-study circumstances, some participants turned to other online platforms to find more relevant forms of support. For example, several participants used KakaoTalk Open Chat groups~\footnote{KakaoTalk is a widely used messenger platform in South Korea, and the Open Chat is a feature that allows users to create chat groups that are open to the public.} organized around graduate student mothers, where they could exchange academic and parenting experiences with similar others. Other participants searched platforms such as Threads, Instagram, and YouTube to find traces of how other graduate student mothers navigated their challenges. Danggeun, a hyper-local community-based secondhand marketplace platform, was also a popular choice among participants for getting practical support by obtaining secondhand childcare products or local information. P22 created a separate community for mothers interested in engaging more deeply with childcare-related learning and recruited members through Instagram. These practices show that participants were not simply disengaging from Mom Cafes, but rather they were actively assembling support across multiple platforms in response to the partial and fragmented support available in any single community.

Taken together, these findings show that Mom Cafes reproduced the invisibility described in the previous section, where participants were visible as mothers within parenting communities, but their identities as graduate students remained difficult to articulate and risky to disclose. As P8 summarized, \textit{``Mom Cafes are needed for moms, not graduate student mothers.''} Many nonetheless continued to use them because the parenting support they provided remained useful.

\subsection{Discussing Safe Visibility: Rethinking Mom Cafes for Graduate Student Mothers}

Building on these experiences, the design workshop invited participants to imagine a Mom Cafe specifically for graduate student mothers. Because this framing was given by the researchers, we do not interpret participants’ designs as evidence that they independently preferred a specialized community. Instead, we use the designs to examine the forms of support they desired and the tensions those forms introduced. Rather than simply requesting new features, participants' ideas surfaced three competing yet interdependent needs: becoming recognizable without becoming individually traceable, developing belonging without incurring social and emotional pressure, and extending online connections into practical support while ensuring that care providers could be trusted. We organize their ideas around these three tensions, which together shaped their visions of \textit{safe visibility}.


\subsubsection{Seeking Recognition While Managing Exposure}


Participants discussed the need for spaces where graduate student motherhood could become recognizable without requiring repeated explanation or justification. In both offline and online settings, they often found it difficult to make their combined academic and caregiving circumstances understandable when those circumstances mattered. They were therefore excited to imagine a space where they could bring this aspect of themselves forward when they chose to do so.

\begin{quote}
    \textit{``I think it would be a space that allows me to be myself. I need a space where I can openly express my own desires --- my desire to study, my desire to grow and develop as a person, and also my desire to be a good mother to my child.''} (P14)
\end{quote}

Participants translated this desire into concrete structures for making similar others findable. They proposed topic-specific boards organized by academic field, degree stage, or child age, contrasting these with existing online communities that they described as \textit{``hectic''} (P19) and difficult to navigate. Seeing how other graduate student mothers had handled comparable situations, including successfully completing graduate study, was a source of both practical guidance and hope.

However, the visibility they desired also raised concerns about exposure. Participants distinguished between being recognized as part of the shared category of graduate student motherhood and being identifiable as a specific individual. Although they wanted their identities acknowledged, they remained cautious about disclosing details such as their university, field of study, advisor relationship, child's age, or residential area. Because they perceived the community of graduate student mothers as small, they believed that combinations of these details could make them recognizable.

This concern also shaped their use of KakaoTalk Open Chat groups for graduate student mothers. While these groups enabled more comfortable sharing and emotional support, participants still managed disclosure carefully because the community remained socially narrow. This discouraged participants from fully disclosing the circumstances surrounding the challenges they experienced as graduate student mothers.

\begin{quote}
    \textit{``I liked the community, so I started recommending it to graduate student mothers around me who were in similar situations. But once you share your department and roughly how old your child is, I feel like people I know could figure out who I am. That seems to reduce the anonymity of the online community and limit how honestly I can speak. And if the community continues to grow and members interact more frequently, both online and offline, I worry that certain impressions or stereotypes may become attached to particular academic fields. That would probably make me more passive in the community.''} (P20)
\end{quote}

Some participants therefore preferred a nationwide community over smaller, locally based ones, while acknowledging that locality was a strength of existing Mom Cafes because it enabled access to neighborhood-specific childcare information (P5, P6, P8). Their discussions revealed a tension between the usefulness of local support and the risk of identification. Participants thus envisioned communities that enabled recognition while allowing members to control how much personal information they disclosed.

\subsubsection{Fostering Belonging While Managing Social and Emotional Burden}

Participants expected a specialized Mom Cafe to foster belonging and empathy among graduate student mothers. They described two forms of belonging. First, simply knowing that others faced similar academic and caregiving demands could alleviate isolation and anxiety, even without direct interaction. The imagined community was therefore not only an information source but also a space that could make the existence of graduate student mothers collectively visible without requiring each member to become individually identifiable. As P17 explained, having a space where she could \textit{``participate and feel a sense of solidarity''} was itself reassuring. Both the visibility of individual cases and the existence of the community as a whole therefore represented potential sources of emotional support.

\begin{quote}
    \textit{``Just reading (during the workshop) that other people were going through the same difficulties I had experienced was comforting. So, if I had to choose the one thing I needed most, I would want a space where we could feel that people like us exist, that we are all living through and managing these everyday realities together.''} (P12)
\end{quote}

Second, participants imagined a community in which members could explicitly comfort and encourage one another. Such active encouragement was often absent from the communities in which they currently participated.

\begin{quote}
    \textit{``I want us to feel that we are not the only graduate student mothers going through these difficulties. I thought that even if each of us seemed rare on our own, there might actually be many of us once we came together. So I imagined a space where, when I shared the struggles I faced as a graduate student mother, I could receive many empathetic reactions and comments. Through those responses, I hoped that my passion for this career path, my commitment to it, and even my sense of calling could be recognized as meaningful. I wanted it to be a space where we could say, `You too? Me too,' support one another, and find comfort when things became difficult.''} (P14)
\end{quote}

Notably, participants did not equate belonging with intimacy. Several preferred loose connections or \textit{``weak ties,''} as P22 described them, that provided reassurance and occasional exchange without reciprocal obligations. This preference also reflected their limited capacity for sustained participation. Some noted that existing Mom Cafes required repeated posting or commenting to maintain membership status or access certain boards, adding participation work to already time-constrained routines (P14, P17). They therefore valued communities where belonging could be sustained without frequent engagement.

Participants also recognized that visibility could become burdensome. Seeing other graduate student mothers succeed academically while raising children could invite comparison and self-imposed pressure. Participants noted that platforms such as Threads and Instagram often showcased highly curated accounts of academic success, making them question whether they were doing enough.

\begin{quote}
    \textit{``Seeing people work so hard on their research while also raising their children (on online communities) can be very helpful and motivating. But when I am not doing well myself, it can also feel like a personal burden. Some graduate student mothers also use Threads as a form of self-promotion, almost like LinkedIn, posting about the conferences they attend or the research they have published. When I see those posts, I sometimes start wondering whether I should be doing more as well, and that can make the platform feel more pressuring than supportive.''} (P21)
\end{quote}

Greater visibility also raised concerns about normative judgment. Having observed mothers being judged in Mom Cafes for departing from dominant expectations of intensive motherhood, participants anticipated that a specialized community would not automatically be exempt from such evaluation. They expected that disclosing academic commitments could still prompt others to question whether they were appropriately prioritizing their children. They therefore emphasized clear community norms and active moderation to create conditions where members could discuss consequential experiences without defending the legitimacy of their choices.

\begin{quote}
    \textit{``In my design, I drew an arrow toward conflict resolution and wrote something like, `Only comment when you have something helpful to say or a possible solution --- no unsolicited opinions or criticism.' I imagined a community centered on unconditional support and practical solutions rather than criticism or personal judgment. I thought that such a community would allow people to find comfort and useful information without becoming emotionally exhausted.''} (P16)
\end{quote}

\subsubsection{Extending Practical Care While Negotiating Trust and Safety}


Finally, participants emphasized that meaningful support should extend beyond online conversation. P17 described this aspiration as \textit{``practical solidarity.''} While emotional reassurance and informational exchange were important, participants also needed practical support for managing academic and caregiving responsibilities.

Participants proposed urgent childcare services organized locally or through universities, as well as cooperative childcare among community members. For example, P20 proposed alternating between research and childcare with other graduate student mothers at 30-minute intervals. Others proposed \textit{``study together''} sessions with other graduate student mothers, particularly through remote video calls or late at night after their children had gone to sleep (P5, P10, P12, P13, P19), to help sustain motivation and concentration despite caregiving and household demands.

However, moving from support through conversation to practical care fundamentally changed what members needed from the community. Participants questioned whether other members were reliable, particularly when children's well-being was involved. Although participants imagined relying on one another, P21 noted that urgent childcare arrangements immediately \textit{``raise concerns about whether the babysitters could be trusted.''}

To manage these risks, participants proposed mechanisms for establishing trust and accountability. Some suggested screening or verifying community members, while others preferred qualified caregivers whose credentials and prior reviews could be inspected. These proposals revealed a tension between the accessibility and reciprocity of peer-based support and the safeguards of formalized services. Participants wanted practical solidarity, but also mechanisms for determining who could provide care safely and responsibly.

Across these tensions, participants did not imagine a specialized Mom Cafe as a simple solution to the limitations of existing parenting communities. Rather, their ideas showed that supporting graduate student mothers requires carefully calibrating visibility: making relevant aspects of their circumstances recognizable when needed without producing unwanted exposure, social comparison, normative judgment, or safety risks. Designing for graduate student mothers therefore involves not only creating a space for similar others to gather, but also mechanisms that allow members to control how much of themselves becomes visible, to whom, and for what forms of support.
\section{Discussion}
In the following sections, we elaborate \textit{safe visibility} by considering how online communities might support recognition without unwanted exposure, connection without excessive relational burden, and practical care with sufficient trust and accountability. Unlike disclosure-centered accounts that primarily concern whether and how much information users reveal, safe visibility foregrounds the ability to make only the circumstances relevant to a particular support need recognizable while independently calibrating identifiability, relational obligation, and task-specific accountability. We first examine how the first two tensions shape participation and belonging in online communities, before turning to what changes when support extends to materially consequential forms of care.

\subsection{Negotiating Recognition and Relational Involvement}


Our findings complicate a simple assumption that greater visibility necessarily leads to better support. Participants wanted relevant aspects of their circumstances to become recognizable when needed, while retaining control over the exposure and relational involvement that followed. Partial recognition was often sufficient when it aligned with the support sought, as in participants’ continued use of Mom Cafes for parenting support. This shifts the design problem from maximizing identity visibility to calibrating which aspects of a person become recognizable for a particular need.

Also, a seemingly straightforward response to the invisibility that participants experienced would be to create specialized communities for people who share particular combinations of circumstances. Such spaces can provide valuable recognition and connection with similar others~\cite{parenting05}. Yet specialization alone is not necessarily the answer. If support for each overlapping circumstance is available only in a specialized community, users may have to seek out increasingly narrow spaces as different needs arise. This can also leave broader communities’ assumptions about their members unchanged. For small populations such as graduate student mothers, narrower membership can also increase the risk of individual identifiability~\cite{ammari19self}.

Our findings instead suggest that broader online communities could become more responsive to the different circumstances their members bring to particular support needs. Broader communities could cultivate norms and structures that allow members to bring relevant aspects of their circumstances forward when needed. The goal is not to make every aspect of a person permanently visible, but to create broader spaces in which different and overlapping circumstances can become recognizable and supported when they matter.

Meanwhile, becoming recognizable raised a second question: what kind of relationship should follow? Participants valued finding others who understood their circumstances, but they did not necessarily want that recognition to develop into sustained relationships or reciprocal obligations. Granovetter’s notion of weak ties~\cite{granovetter1977strength} offers a useful starting point for understanding this preference for loose connections in online communities. Our findings, however, suggest that the value of such ties extended beyond access to diverse information. For graduate student mothers, low-obligation ties could reduce the relational costs of support by enabling recognition and solidarity without requiring sustained self-disclosure, reciprocal engagement, or relationship maintenance. This was particularly consequential for participants already facing acute time scarcity across research and caregiving, for whom denser ties could introduce additional burdens through social comparison, normative judgment, and expectations of reciprocity.

Seen this way, the fragmented support ecology described in our findings need not be understood only as a deficiency to be repaired. This fragmented support ecology echoes prior research showing that people may turn to separate online networks when particular experiences are difficult to discuss within existing networks~\cite{zhang22separate, zhao16social, boc18plat}. We extend this perspective by showing that such distribution can also function as a way of managing relational exposure and obligation in support-seeking. By moving across Mom Cafes, KakaoTalk Open Chats, and other social platforms, they could seek support in different contexts without becoming deeply embedded in any one of them. The design challenge, then, may not be to consolidate this ecology into a single community, but to support movement across different contexts while preserving members’ control over how much relational commitment each interaction entails.

These findings show that safe visibility concerns not only what becomes visible, but also what relational demands follow from that visibility. Supporting people in becoming recognizable to relevant others does not require making more of the self permanently visible, nor does recognition need to develop into sustained social relationships. Relational dialectics conceptualizes apparently opposing relational needs not as problems to be permanently resolved, but as tensions that are continually negotiated in relation to context~\cite{baxter1996relating, kim07cy}. Our findings reveal a similar dynamic in online support: participants sought recognition without unrestricted exposure and connection without sustained relational obligation. Building on this perspective, safe visibility foregrounds how sociotechnical environments should shape people’s ability to recalibrate these tensions according to the support at stake. 


\subsection{Connecting Online Parenting Communities to Practical Care Infrastructures}
Low-obligation connection becomes difficult to sustain when support moves from recognition and conversation to materially consequential care. The limits of online social support became particularly visible when participants needed not information or empathy, but time and care capacity. Advice could help participants respond when childcare arrangements failed, and emotional support could make such experiences feel less isolating, but neither could create the time needed to conduct research, prepare for a dissertation defense, or meet a deadline while caring for a child. Care thus extended beyond childcare to sustaining mothers’ capacity to remain students and researchers while fulfilling caregiving responsibilities. Their designs therefore point toward an infrastructure of mutual care: not simply an online community that understands graduate student motherhood, but a sociotechnical arrangement that helps preserve the material conditions necessary for inhabiting both academic and caregiving roles.

Prior work conceptualizes care as infrastructure and shows both how digital spaces can support informal care and their limits in substituting for material childcare~\cite{li23care}. HCI and CSCW research likewise shows that digital platforms can mediate care beyond the platform~\cite{nanth2026}, while also generating expectations of availability and relational labor~\cite{dye2021}. Our findings extend this work by showing that moving from online support to materially consequential care changes what safety and trust need to accomplish. When participants merely exchanged information or emotional reassurance, relatively anonymous or lightweight interactions could be sufficient. Childcare is different because accepting support requires entrusting another person not only with a child’s well-being, but also with a scarce period of time on which one’s academic responsibilities may depend. A failed childcare arrangement may therefore produce consequences across both roles: concern for the child as well as the loss of the time that the arrangement was intended to make available for research. The relevant requirement is not simply interpersonal trust, but reliability and accountability sufficient for members to depend on care without creating another source of uncertainty or repair work. This helps explain why participants proposed mechanisms for evaluating whether care providers could act responsibly, such as verification or reviews.

Yet this requirement conflicts with the low-obligation connections participants valued. Building trust for childcare may require repeated interaction, greater disclosure, and stronger ties, but these same processes can reintroduce risks of identifiability and judgment, as well as reciprocity and relationship-maintenance burdens. Dye’s account of infrastructural care illustrates a similar paradox: communication technologies enabled care across distance, but the possibility of constant connection also generated expectations of constant availability and increased relational labor~\cite{dye2021}. For graduate student mothers already experiencing severe time scarcity, a care infrastructure that reduces childcare labor while requiring substantial new relationship-maintenance labor may simply relocate rather than reduce their burden.

Thus higher-stakes support need not necessarily require turning weak ties into strong ties. Instead, online communities could separate accountability from relational intensity. Connections could remain bounded and low-obligation while providing enough accountability for particular practical tasks. For example, members could interact pseudonymously by default and verify additional information only when arranging childcare. Reliability might also be supported through task-specific histories or credentials, without requiring members to know one another well. Communities could further reduce the burden on individual mothers by connecting them to university childcare resources, vetted professionals, or institutionally supported peer-care pools. The success of such systems should therefore be evaluated not by whether they produce more interaction or denser relationships, but by whether they allow members to mobilize care reliably while preserving their ability to decide how much relational involvement that care should require.

\section{Limitations and Future Work}

This study examines graduate student mothers’ experiences with Mom Cafes in South Korea. The social meanings of graduate study and motherhood, and institutional conditions around childcare, parental support, and graduate education, vary across contexts. Accordingly, how selective visibility is negotiated, and the tensions among recognition, identifiability, relational obligation, and trust, may manifest differently elsewhere. Here, cross-cultural studies could examine graduate student mothers situated in different academic and caregiving contexts to understand which aspects of safe visibility generalize across settings and which are shaped by local norms and institutional arrangements. Future studies could also examine other types of online parenting communities, including local community platforms, social media, and messenger-based groups, to understand how different platform structures shape members' experiences of seeking support across intersecting identities.

Our design workshops were also intentionally framed around participants' existing experiences with Mom Cafes. Although this grounding helped participants articulate concrete needs and tensions, asking them to envision a Mom Cafe for graduate student mothers may have anchored their ideas to familiar community structures and constrained more radical alternatives. Moreover, the workshops focused primarily on envisioning desirable communities rather than designing and evaluating implementable features for online communities from scratch for graduate student mothers. Future work could therefore engage graduate student mothers in iterative participatory design to develop and evaluate ways of finding relevant others while preserving control over visibility and keeping participation low-obligation, especially where task-specific trust is needed. Such work could examine how higher-stakes practical support can be enabled while preserving members’ control over identity disclosure, identifiability, and relational involvement.
\section{Conclusion}

This study examined how graduate student mothers navigate online support while balancing graduate study and caregiving. Participants experienced dual invisibility, where they were recognized as graduate students in academic settings and as mothers in parenting settings, while the overlap between these roles often remained difficult to make legible when support needs depended on both. Mom Cafes remained valuable for parenting information and emotional reassurance, but concerns arising specifically from graduate student motherhood were often harder to articulate or safely disclose. Importantly, our findings suggest that support does not require users to make their full identities persistently visible. Rather, participants valued the ability to make relevant aspects of their circumstances recognizable when needed while retaining control over what they disclosed, how identifiable they became, and what relational obligations followed. We conceptualize these tensions as safe visibility. Thus, we suggest that online support infrastructures support selective, context-dependent recognition and low-obligation connection while enabling task-appropriate trust and accountability for practical care.


\bibliographystyle{ACM-Reference-Format}
\bibliography{Main}

\appendix
\clearpage
\appendix

\section{Interview Protocol}
\label{app:inter}
This section presents the set of interview questions used during the study. As the interviews were semi-structured, the exact wording and order of the questions varied across participants depending on the flow of each interview.

\subsection{Interview for Mom Cafe Active Users}
\subsubsection{Academic and Parenting Background}
Please briefly introduce yourself for about one minute. What is your current degree status (master’s student, doctoral student, coursework completed, or graduate), and what is your field of study?
\begin{itemize}
    \item Did you enter graduate school immediately after completing your undergraduate degree, or did you work before enrolling?
    \item At what point during your graduate studies did you give birth and begin raising your child?
\end{itemize}

\subsubsection{Challenges of Childbirth and Parenting During Graduate School}
What were the greatest challenges you faced, or the areas in which you needed the most support, while experiencing childbirth and parenting as a graduate student?
\begin{itemize}
    \item Were there any challenges that directly affected your academic work or research?
    \item Compared with mothers who are not attending graduate school, such as working mothers or stay-at-home mothers, were there any challenges that you found particularly difficult?
    \item What strategies did you use to address these challenges? Which strategy was the most helpful?
\end{itemize}

\subsubsection{Initial Motivation for using Mom Cafe}
What initially led you to use a Mom Cafe?

\subsubsection{Purposes and Patterns of Mom Cafe use}
What do you primarily use Mom Cafes for?
\begin{itemize}
    \item What kinds of posts or comments do you usually read or write?
\end{itemize}

\subsubsection{Offline Relationships Through Mom Cafe}
Have you ever met someone in person whom you first encountered through a Mom Cafe, or participated in any offline activities through connections formed there?
\begin{itemize}
    \item Could you describe the experience in detail?
\end{itemize}

\subsubsection{Comparison Between Online and Offline Communication About Graduate Student Motherhood}
How does communicating through a Mom Cafe differ from talking offline with friends, family members, or other graduate students about your life as a graduate student and mother?

\subsubsection{Helpful Experiences in Mom Cafe}
Could you describe a specific situation in which a Mom Cafe was helpful to you?
\begin{itemize}
    \item Have you ever used a Mom Cafe to address challenges you faced as a graduate student?
    \item After receiving help from other members, have you ever provided help to other members in return?
\end{itemize}

\subsubsection{Negative Experiences in Mom Cafe}
Conversely, have you had any experiences in which using a Mom Cafe negatively affected your actual parenting practices or experiences, or in which you were unable to obtain the help you wanted?

\subsubsection{Limitations of Mom Cafe for Graduate Student Mothers}
Do you feel that other members of the Mom Cafe understand your circumstances as a graduate student? Why or why not?
\begin{itemize}
    \item Are there aspects of the challenges faced by graduate student mothers that Mom Cafes are not well equipped to address?
    \item What kinds of issues do you think are not adequately addressed?
\end{itemize}

\subsubsection{Use of Other Online Parenting Communities and Platforms}
Do you use any other online communities or platforms for parenting, such as KakaoTalk Open Chat, blogs, or Instagram?
\begin{itemize}
    \item What initially led you to use these communities or platforms, and why do you continue to use them?
    \item Compared with Mom Cafes, why did you decide to use these communities or platforms alongside Mom Cafes?
\end{itemize}

\subsubsection{Future Need for Online Parenting Communities Among Graduate Student Mothers}
Do you think online communities such as Mom Cafes will continue to be necessary for graduate student mothers? Why or why not?

\subsection{Interview for Mom Cafe Non-Active Users}
\subsubsection{Academic and Parenting Background}
Please briefly introduce yourself for about one minute. What is your current degree status (master’s student, doctoral student, coursework completed, or graduate), and what is your field of study?
\begin{itemize}
    \item Did you enter graduate school immediately after completing your undergraduate degree, or did you work before enrolling?
    \item At what point during your graduate studies did you give birth and begin raising your child?
\end{itemize}

\subsubsection{Challenges of Childbirth and Parenting During Graduate School}
What were the greatest challenges you faced, or the areas in which you needed the most support, while experiencing childbirth and parenting as a graduate student?
\begin{itemize}
    \item Were there any challenges that directly affected your academic work or research?
    \item Compared with mothers who are not attending graduate school, such as working mothers or stay-at-home mothers, were there any challenges that you found particularly difficult?
    \item What strategies did you use to address these challenges? Which strategy was the most helpful?
\end{itemize}

\subsubsection{Reason for Not Using Mom Cafe}
Why have you chosen not to use Mom Cafes?
\begin{itemize}
    \item Do you plan to continue not using Mom Cafes? Why or why not?
\end{itemize}

\subsubsection{Use of Alternative Online Parenting Communities and Platforms}
Do you use any online communities or platforms for parenting, such as KakaoTalk Open Chat, blogs, or Instagram?
\begin{itemize}
    \item What initially led you to use these communities or platforms, and why do you continue to use them?
    \item Compared with Mom Cafes, why did you choose to use these communities or platforms instead?
\end{itemize}

\subsubsection{Purposes and Patterns of Alternative Platform Use}
What do you use these communities or platforms for? What kinds of posts do you tend to read or write, and what kinds of posts do you comment on?
\begin{itemize}
    \item How does communicating through these communities or platforms differ from talking offline with friends, family members, or other graduate students about your life as a mother?
\end{itemize}

\subsubsection{Experiences of Alternative Platforms}
Could you describe a specific situation in which these communities or platforms were helpful to you?
\begin{itemize}
    \item Have you ever used these communities or platforms to address challenges you faced as a graduate student?
    \item After receiving help from other members, have you ever provided help to other members in return?
    \begin{itemize}
        \item What kind of help did you provide, and why did you decide to help?
    \end{itemize}
    \item Conversely, have you had any experiences in which using these communities or platforms negatively affected your actual parenting practices or experiences?
    \item Are there aspects of the challenges faced by graduate student mothers that these communities or platforms are not well equipped to address?
    \begin{itemize}
        \item What kinds of issues do you think are not adequately addressed?
    \end{itemize}
\end{itemize}

\subsubsection{Reasons for Not Using Any Online Parenting Communities}
Is there a particular reason why you decided not to use any online communities or platforms for parenting? If so, what is it?

\subsubsection{Alternative Sources of Parenting Information and Support}
Online communities such as Mom Cafes are known to provide a wide range of parenting-related and local information. How do you obtain this type of information or support?

\subsubsection{Future Need for Online Parenting Communities Among Graduate Student Mothers}
Do you think online communities such as Mom Cafes will continue to be necessary for graduate student mothers? Why or why not?





\clearpage
\section{Challenges of Graduate Student Mothers}
\label{app:gsmchall}
This section summarizes the challenges of graduate student mothers identified from participants’ accounts. These challenges provide broader context for the identity ~---~ and support ~---~ related experiences examined in the main findings. For each challenge, we provide an illustrative example based on specific experiences shared by participants during \textit{Part 1} of the workshop.

\begin{table}[htbp]
\caption{Challenges experienced by graduate student mothers.}
\Description{A three-column table presenting 17 challenges experienced by graduate student mothers, organized into seven categories: physical, career, economic, emotional, social, instrumental, and overlapping responsibilities. The columns identify each category, a specific challenge within that category, and an illustrative example from the design workshop. Examples include pregnancy- and childbirth-related difficulties, disrupted research and career opportunities, financial insecurity, guilt and isolation, inadequate childcare support, and conflicts between academic and caregiving responsibilities.}
\label{tab:challenge}

\small
\setlength{\tabcolsep}{4pt}
\renewcommand{\arraystretch}{1.08}

\begin{tabularx}{\linewidth}{
    @{}
    >{\raggedright\arraybackslash}p{0.12\linewidth}
    >{\raggedright\arraybackslash}p{0.30\linewidth}
    >{\raggedright\arraybackslash}X
    @{}
}
\toprule
\textbf{Category} &
\textbf{Challenge} &
\textbf{Examples from Design Workshop Part 1} \\
\midrule

Physical
& Physical challenges associated with pregnancy
& Difficulty concentrating on advisor-assigned research due to physical exhaustion during pregnancy. \\

& Physical challenges associated with childbirth
& Experiencing reduced mental sharpness and memory after childbirth. \\

\midrule

Career
& Limited access to career development opportunities
& Losing access to academic information and opportunities while on post-childbirth leave. \\

& Difficulty sustaining research due to childcare
& Pausing an ongoing project around childbirth and starting anew when resuming it proved difficult. \\

& Children in career decision-making
& Having limited flexibility in post-graduate career decisions because of childcare responsibilities. \\

& Research collaboration barriers due to childcare
& Being told by an advisor that frequent childcare-related lab absences made collaboration difficult. \\

\midrule

Economic
& Financial insecurity
& Being unable to plan long-term educational activities for a child because of unstable income. \\

\midrule

Emotional
& Psychological burden from multiple responsibilities
& Meeting academic demands despite inflexible schedules and reluctance to cite childcare as a constraint. \\

& Sense of guilt toward child
& Feeling guilty when a child wakes during late-night work and says, ``Mom, don't study.'' \\

& Lack of peers with shared experiences
& Lacking peers who could relate to pregnancy, childbirth, and childcare-related struggles. \\

& Comparison with other graduate students
& Feeling envious of peers who could focus on papers without childcare responsibilities. \\

\midrule

Social
& Graduate study perceived as a selfish choice for mothers
& Facing in-law criticism for pursuing a PhD beyond a master's degree. \\

& Lower perceived value of graduate study than paid work
& Encountering the view that graduate study does not count as ``working.'' \\

\midrule

Instrumental
& Lack of childcare-friendly spaces on campus
& Struggling to find adequate time and space for pumping breast milk on campus. \\

& Lack of support for childcare-related leave
& Returning to university when childcare leave ended despite having no available caregiver. \\

\midrule

Overlap of responsibilities
& Pressure to relinquish responsibilities
& Missing a paper submission deadline because of a child's unexpected hospitalization. \\

& Challenges in time management
& Repeatedly interrupting research for childcare and retracing prior work after each interruption. \\

\bottomrule
\end{tabularx}
\end{table}

\clearpage
\section{Workshop Results}
\label{app:workshop}
This section presents selected excerpts and examples of sketches created by participants during \textit{Part 2} of the workshop. The sketches illustrate participants’ design ideas and the support needs and tensions that directly relate to our findings. Blue text indicates English translations of the original Korean annotations.

\begin{figure}[htbp]
    \centering
    \includegraphics[width=\linewidth]{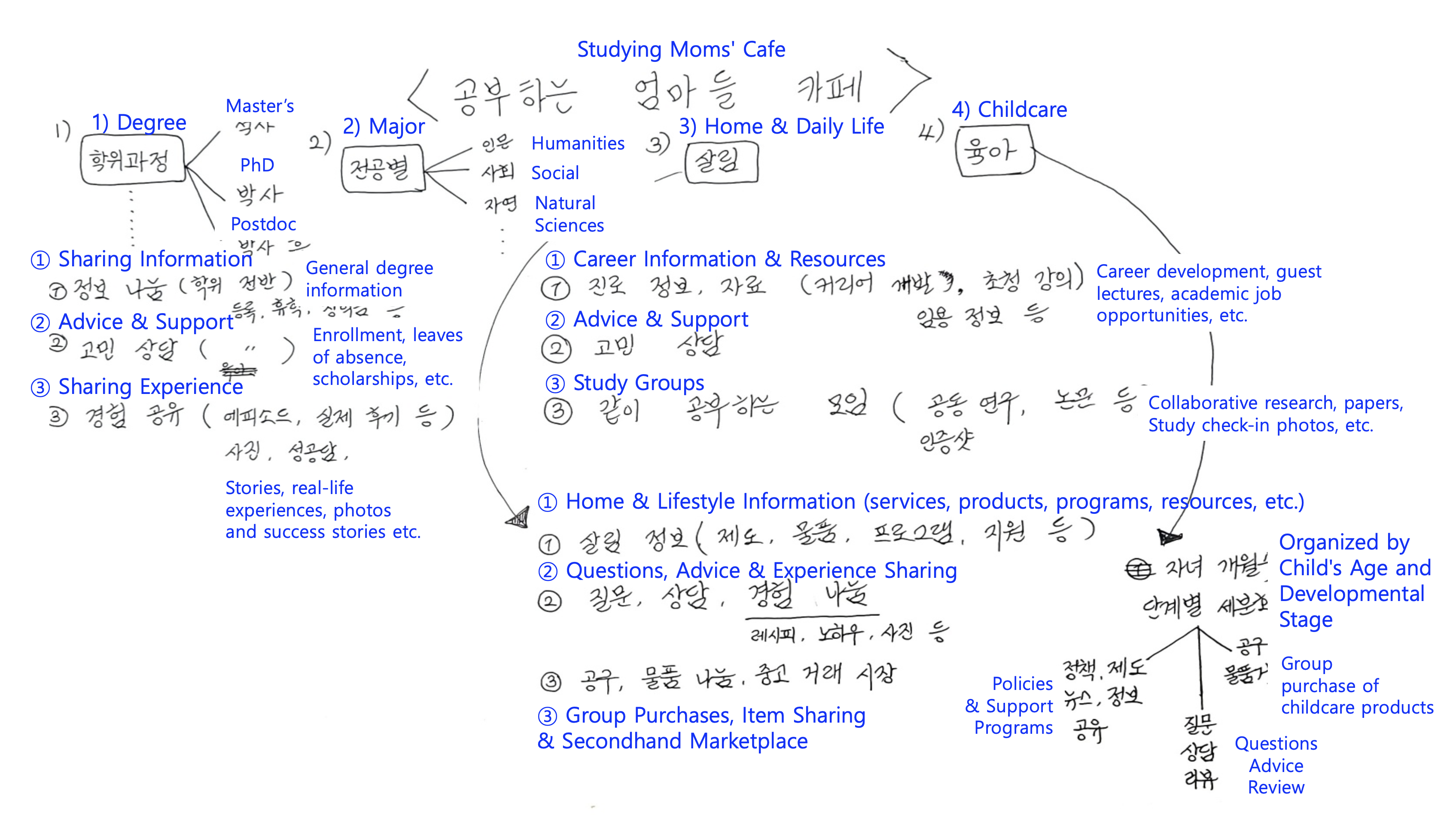}
    \caption{A sketch of a Mom Cafe for graduate student mothers where they can easily find similar others for informational and emotional support between community members. }
    \Description{Participant sketch of a ``Studying Moms Cafe'' organized as a branching community structure. Members can navigate spaces based on characteristics such as degree, academic field, home and daily life, and children. Subsections support sharing information and experiences, seeking advice and emotional support, finding career and academic resources, forming study groups, exchanging local and household information, and connecting around children's ages and developmental stages.}
    \label{fig:empathy-workshop}
\end{figure}

\begin{figure}[htbp]
    \centering
    \includegraphics[width=\linewidth]{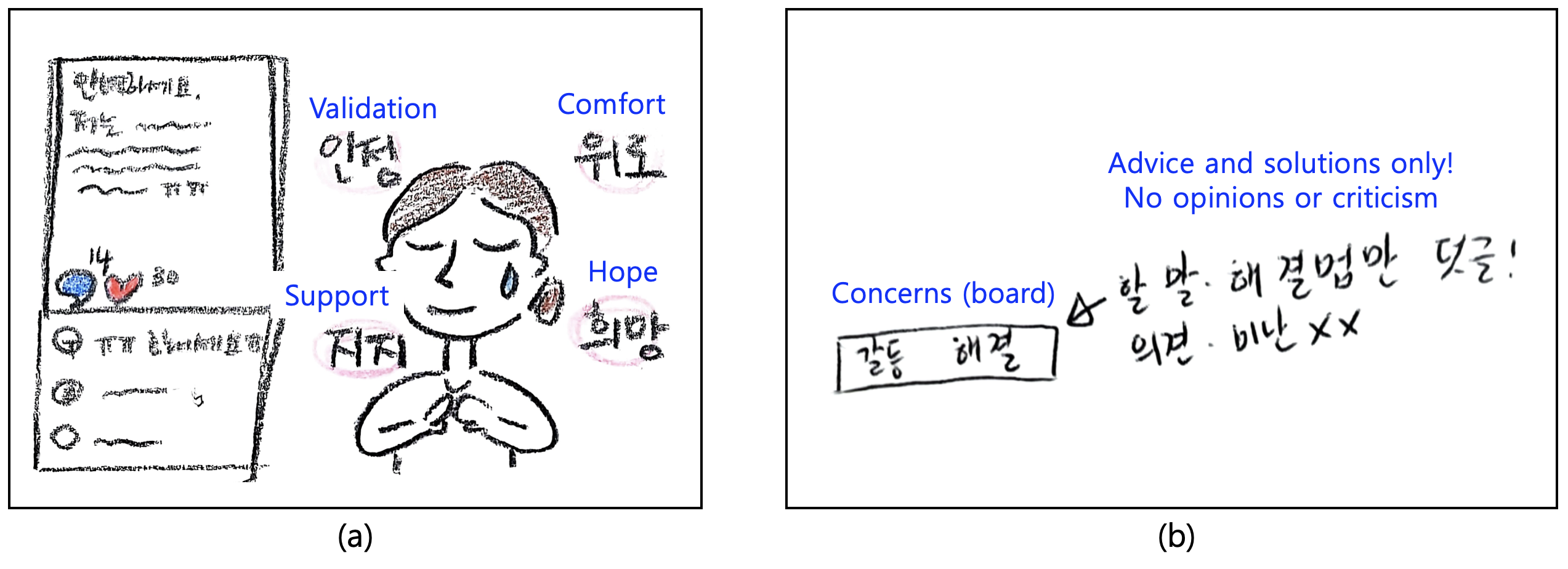}
    \caption{Participants’ design ideas for supportive online interactions: (a) Fostering validation, comfort, support, and hope between members; (b) A board where members can share concerns, while encouraging advice and solutions without opinions or criticism.}
    \Description{Two participant sketches illustrating emotionally supportive interaction. Panel (a) shows a community member surrounded by four desired forms of response: validation, comfort, support, and hope. Panel (b) depicts a concerns board governed by a norm of providing helpful advice and solutions while discouraging unsolicited opinions or criticism.}
    \label{fig:tension2}
\end{figure}

\begin{figure}[htbp]
    \centering
    \includegraphics[width=\linewidth]{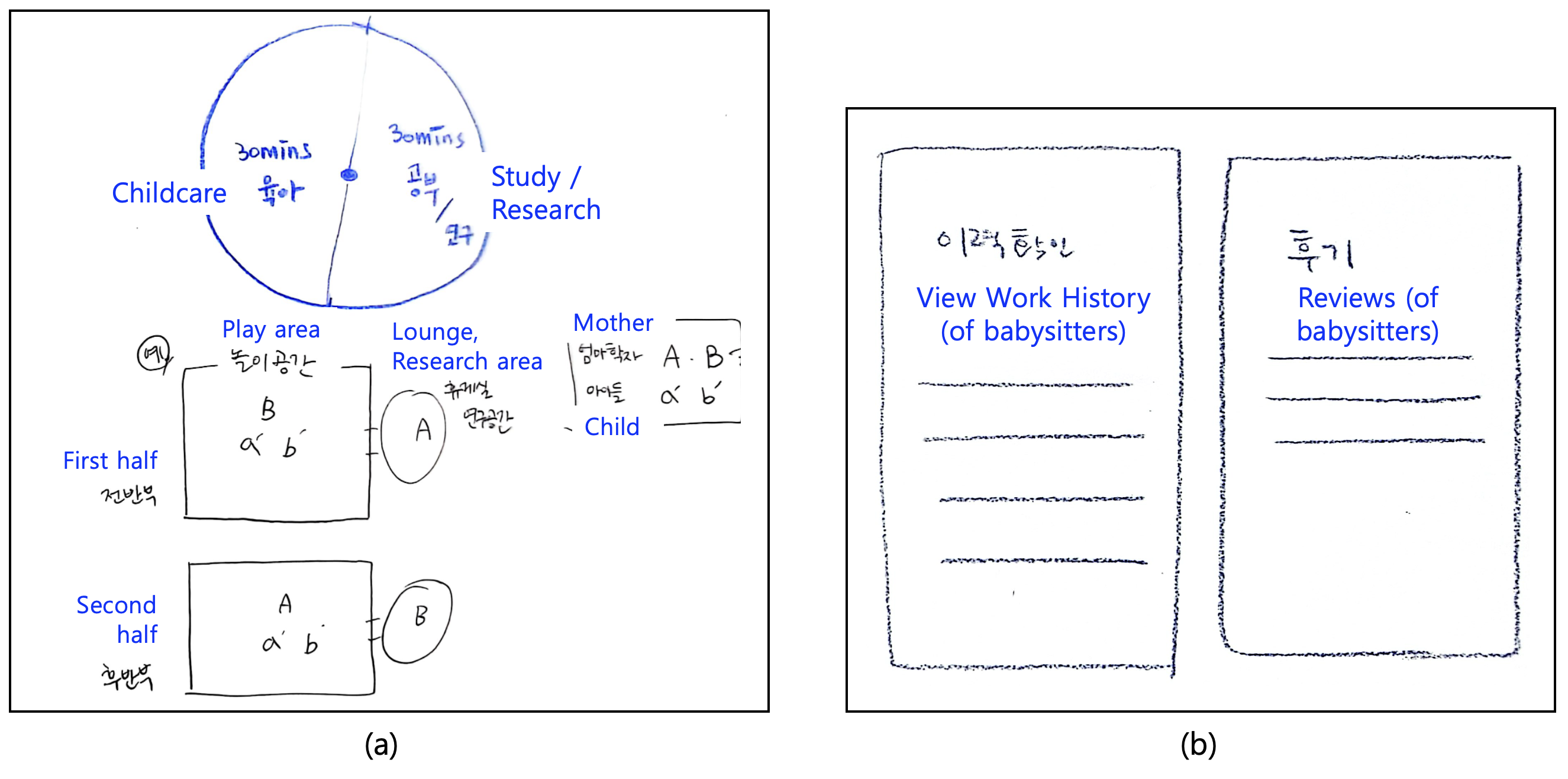}
    \caption{Participants’ proposed ways of supporting parents in balancing childcare and academic work: (a) Organizing childcare and study or research activities by switching every 30 minutes with other graduate student mothers through cooperative childcare; (b) Providing babysitters’ work histories and reviews to make a safer caregiver selection.}
    \Description{Two participant sketches of practical support mechanisms. Panel (a) depicts two graduate student mothers alternating childcare and study or research in 30-minute intervals, so that one cares for the children while the other works and then they switch roles. Panel (b) depicts a caregiver-selection interface displaying babysitters' work histories and reviews to support safer and more accountable childcare decisions.}
    \label{fig:tension3}
\end{figure}

\end{document}